\documentclass{article}
\usepackage{spconf,graphicx}
\usepackage[utf8]{inputenc}
\usepackage[T1]{fontenc}
\usepackage[american]{babel}
\usepackage{graphicx}
\usepackage{cite}
\usepackage{listings}
\usepackage{float}
\usepackage{amsmath,amssymb,exscale}
\usepackage{blindtext, graphicx}
\usepackage{verbatim}
\usepackage{algorithm}
\usepackage{algpseudocode}
\usepackage{fancyvrb}
\usepackage{bera}
\usepackage{mathtools}
\usepackage{lipsum}

\usepackage{times}
\usepackage{scalerel}
\usepackage{tikz}
\usetikzlibrary{svg.path}
\usepackage[bookmarks=false]{hyperref}

\title{ Optimal BER Minimum Precoder Design for OTFS-Based ISAC Systems}

\name{Jun Wu\textsuperscript{1},
     Weijie Yuan\textsuperscript{1},
      Zhiqiang Wei\textsuperscript{2},
       Jinjin Yan\textsuperscript{3},
    \text{and} Derrick Wing Kwan Ng\textsuperscript{4}
}

 \address{\textsuperscript{1} Department of Electrical and Electronic
Engineering, \\ Southern University of Science and Technology, Shenzhen, China \\
         \textsuperscript{2} School of Mathematics and Statistics, Xi’an Jiaotong University, Xi'an, China \\
         \textsuperscript{3} Innovation and Development Center, Harbin Engineering University, Qingdao, China \\
         \textsuperscript{4} School of Electrical Engineering and Telecommunications, \\ University of New South Wales, Sydney, Australia \\     
         (\textit{Invited Paper})
}
     
\begin{document}
%
\maketitle
\begin{abstract}
This paper investigates the bit error rate (BER) minimum precoder design for an orthogonal time frequency space (OTFS)-based integrated sensing and communications (ISAC) system, which is considered as a promising technique for enabling future wireless networks. In particular, the BER minimum problem takes into account the maximized available transmission power and the required sensing performance.  We devise the precoder from the perspective of delay-Doppler (DD) domain by exploiting the equivalent DD channel. To address the non-convex design problem,  we resort to minimizing the lower bound of the derived average BER. Afterwards, we propose a computationally iterative method to solve the dual problem at low cost. Simulation results verify the effectiveness of our proposed precoder and reveal the interplay between sensing and communication for dual-functional precoder design.
\end{abstract}
\vspace{-0.5cm}
\section{Introduction}
\label{sec:intro}

Integrated sensing and communication (ISAC) has attracted significant attention due to the fact that both sensing and communication functionalities are highly expected in the emerging next-generation wireless networks \cite{9737357}. The so-called ISAC technology refers to the joint design of sensing and communications in a unified device exploiting the same resources and eventually pursuing mutual benefits. By doing so, the system can achieve higher utilization efficiency while saving the cost of spectral resources and power. 
Recently, various interesting schemes have been studied for realizing practical ISAC systems including the joint waveform design and resource allocation \cite{10147248,10005137,10158322}. Meanwhile, motivated by decades of research on orthogonal frequency division multiplexing (OFDM)-based communications, it is a natural progression to design ISAC systems by exploiting OFDM waveforms.
Although OFDM shows its superiority in robust synchronization and resilience against multi-path fading, it suffers from a high peak-to-average-power ratio (PAPR) and has seriously degraded performance due to the severe Doppler spread in high-mobility scenarios. To circumvent this challenge, there is a trend to seek a new modulation scheme, e.g., orthogonal time frequency space (OTFS) \cite{zhang2022deep,9508932,10118784,9940260,zhang2023radar}, which modulates data symbols in the delay-Doppler (DD) domain, different from OFDM that modulates in the time-frequency (TF) domain.  In particular, by exploiting the quasi-static and sparse channel in the DD domain, OTFS is capable of supporting multi-path and high-mobility communications. Furthermore, OTFS modulation directly links the transmitted signals and the DD domain channel characteristics, e.g., delay and Doppler shifts, enabling OTFS as a promising candidate for realizing ISAC \cite{9903393}. 
As such, various works, e.g., \cite{9724198,isac}  have studied radar-assisted communication systems adopting OTFS modulation for both uplink and downlink transmissions. However, these contributions mainly focus on signal processing in the TF domain, which does not fully exploit the the characteristics of DD domain. Meanwhile, The OTFS-based ISAC system design oriented DD domain has not been reported, yet.

Motivated by the above discussions, in this paper, we devise the optimal precoder for OTFS-based ISAC systems in which the bit error rate (BER) and CRB are employed to evaluate the communication performance and sensing accuracy, respectively. We first derive the unified BER expressions for both the cases adopting zero-forcing (ZF) and minimum mean square error (MMSE) equalizations as well as the CRB with respect to (w.r.t) Doppler estimation. Then, the BER minimization problem is formulated subject to the maximum transmit power constraint and required CRB threshold. To address the problem at hand, we rely on the minimization of the average BER lower bound at moderate-to-high SNRs. By leveraging the singular value decomposition (SVD), we derive the structure of optimal precoder based on the Karush-Kuhn-Tucker (KKT) conditions. As such, the minimum BER precoder is obtained by solving the dual problem adopting the gradient descent algorithm. Our simulation results show that our proposed precoder can achieve a considerable BER performance compared with the benchmarks. 

\textit{Notations:} The boldface lowercase letter and boldface capital letter denote the vector and the matrix, respectively. The superscript $(\cdot)^{T}$ and $ (\cdot)^{H}$ denote the transposition, Hermitian operations, respectively.  We adopt $|\cdot|$, $\mathbb{E}(\cdot)$, and tr($\cdot$) to denote the modulus of a complex variable, the expectation operation, and the trace operation, respectively;  $\otimes$, $\mathbf{F}_{\mathrm{M}}$, and $\mathbf{I}_\mathrm{M}$ represent the Kronecker product, the $M$-dimensional discrete Fourier transform (DFT) matrix and identity matrix, respectively.  erfc ($\cdot$) is the complementary error function and we define $[x]^+\triangleq \max(0,x)$. $\mathbb{C}^{M\times N}$ denotes the set of all $M \times N$ matrices with complex entries and $\mathbf{A}_{m,m}$ is the $(m,m)$-th entry of $\mathbf{A}$.  The Gaussian distribution with mean $\boldsymbol{\mu}$ and covariance matrix $\mathbf{\Sigma}$ is denoted by $\mathcal{CN} (\boldsymbol{\mu}, \mathbf{\Sigma})$. $\mathrm{circ}(\cdot), \mathrm{diag}(\cdot)$ represent a  cyclic shift matrix and diagonal matrix, respectively.

\vspace{-0.3cm}
\section{OTFS Modulation and Demodulation}
Let  $\mathbf{x}\in \mathbb{C}^{MN\times 1}$ be the transmitted ISAC symbol vector in the DD domain, where $M$ and $N$ are the number of subcarriers and time slots of each transmission frame,  respectively. Assume that the duration of each time slot is $T$ and the subcarrier spacing is $\Delta f$.   
The vector $\mathbf{x}$ follows that $ \mathbf{x}=\mathbf{W d}$,
where $\mathbf{W} \in\mathbb{C}^{MN \times MN}$ represents the dual-functional precoder matrix to be designed and  $\mathbf{d}\in \mathbb{C}^{MN \times 1}$ is the unit-power information vector generated by the quadrature amplitude modulation (QAM), i.e., $\mathbb{E}\left\{ \mathbf{dd}^H \right\}=\mathbf{I}_{MN}$.
Then, the received signal in the DD domain is given by
\begin{align}
\mathbf{y}= \underbrace{\left(\mathbf{F}_{{N}} \otimes \mathbf{I}_{{M}}\right) \mathbf{H}_\mathrm{T}\left(\mathbf{F}_{{N}}^{{H}} \otimes \mathbf{I}_{{M}}\right)}_{\mathbf{H}_{\mathrm{dd}}} \mathbf{x}+\left(\mathbf{F}_{{N}} \otimes \mathbf{I}_{{M}}\right) \mathbf{n}, \label{hdd}
\end{align}
in which $\mathbf{H}_{\mathrm{dd}} \in \mathbb{C}^{MN\times MN}$ denotes the equivalent channel matrix,  $\mathbf{H}_\mathrm{T}$ is the time domain channel matrix after removing cyclic prefix (CP), and $\mathbf{n}$ is the additive white Gaussian noise (AWGN) vector.  The matrix $\mathbf{H}_\mathrm{T}$ can be characterized by 
 \begin{align}
     \mathbf{H}_\mathrm{T}=\sum_{p=1}^P h_p \boldsymbol{\Pi}^{l_p} \boldsymbol{\Delta}^{k_p}, \label{timechannel}
 \end{align} 
where $P$ and $h_p$ denote the number of paths and complex path gain of the $p$-th path, $l_p=M \Delta f \tau_p$ and $k_p=N T \nu_p$ are the delay and Doppler-shift index, respectively. Here,  $\tau_p$ and $\nu_p$ are the delay and Doppler shift associated with the $p$-th path.
 $\mathbf{\Pi}=\operatorname{circ}\left\{\left[\begin{array}{lll}0 & 1 & 0 \cdots 0\end{array}\right]_{M N \times 1}^{\mathbf{T}}\right\}$ represents the forward cyclic shift matrix, and $\boldsymbol{\Delta}^{k_p} \triangleq$ $\operatorname{diag}\left(e^{\frac{j 2 \pi k_p \times 0}{M N}}, e^{\frac{j 2 \pi k_p \times 1}{M N}}, \ldots, e^{\frac{j 2 \pi k_p \times(M N-1)}{M N}}\right)$.

\vspace{-0.4cm}
\section{Problem Formulation}
\label{sec:pagestyle}
We consider an OTFS-enabled ISAC system, which comprises an ISAC transmitter equipped with one single antenna, a single-antenna communication user, and an unknown target. By transmitting the dual functional signal to the communication user,  the received signal is given by  $ \mathbf{y}_c=\mathbf{H}_{\mathrm{c,dd}} \mathbf{Wd}+ \mathbf{n}_c,$
where $\mathbf{H}_{\mathrm{c,dd}}$ is the equivalent channel to the communication user by modifying $\mathbf{H}_\mathrm{dd}$ defined in (\ref{hdd}) and $\mathbf{n}_c \sim \mathcal{CN}(\mathbf{0},\sigma_c^2\mathbf{I}_{MN})$.
For each received OTFS block, by exploiting the popular ZF/MMSE equalizer $\mathbf{Q}$ represented by {\small $\mathbf{Q}=\left(\kappa\sigma_c^{2}\mathbf{I}_{MN}+\mathbf{W}^{H}\mathbf{H}_{\mathrm{c,dd}}^{{H}}\mathbf{H}_{\mathrm{c,dd}}\mathbf{W}\right)^{-1}\mathbf{W}^{{H}}\mathbf{H}_{\mathrm{c,dd}}^{{H}}$},
where $\kappa$ is an indicator variable, i.e., $\kappa=1$ for MMSE equalizer and $\kappa=0$ for ZF equalizer,
the equalized symbols $ \hat{\mathbf{d}}$ can be recovered as $\hat{\mathbf{d}}=\mathbf{QH}_{\mathrm{c,dd}}\mathbf{W}\mathbf{d}+\mathbf{Q}\mathbf{n}_c.$
Consequently, the received signal-to-interference-plus-noise ratio (SINR) of the $m$ -th ($m\in [1,MN]$) symbol is \cite{9775701}
{\small \begin{align}
\mathrm{SINR}_m =\frac{1}{\left[\left(\kappa\mathbf{I}_{MN}+ \frac{1}{\sigma_c^2}\mathbf{W}^{{H}}\mathbf{H}_{\mathrm{c,dd}}^{{H}}\mathbf{H}_{\mathrm{c,dd}}\mathbf{W}\right)^{-1}\right]_{m,m}}-\kappa. 
\end{align}}
Under the assumption of  Gray encoding,  we can characterize the BER of the $m$-th symbol in the case of  $M_{\mathrm{Mod}}$-ary QAM constellation as $ \mathrm{BER}_m\approx \alpha \mathrm{erfc}\left(\sqrt{\beta \mathrm{SINR}_m}\right),$
where $\alpha=(2-2/\sqrt{M_{\mathrm{Mod}}})/\log_2M_{\mathrm{Mod}}$ and $\beta=3/(2M_{\mathrm{Mod}}-2)$. Thus, the average BER is given by  (\ref{ber}) and shown on the top of next page. 
\begin{figure*}[t]
 {\small    \begin{align}
P_e =\frac{\alpha}{MN} \sum_{m=1}^{MN} \operatorname{erfc}\left(\sqrt{\frac{\beta}{\sigma_c^2\left[\left(\kappa \sigma_c^2 \mathbf{I}_{MN}+\mathbf{W}^{{H}} \mathbf{H}_{\mathrm{c,dd}}^{{H}} \mathbf{H}_{\mathrm{c,dd}} \mathbf{W}\right)^{-1}\right]_{m,m}}-\beta \kappa}\right). \label{ber}
\end{align}}  \hrulefill
\end{figure*}
On the other hand, for sensing the target, the  reflected echo received at the ISAC transmitter is  $ \mathbf{y}_s=\mathbf{H}_\mathrm{s,dd}\mathbf{x}+\mathbf{n}_s,$
where $\mathbf{H}_\mathrm{s,dd}$ is the target response matrix in DD domain and $\mathbf{n}_s$ is the AWGN obeying $\mathbf{n}_s \sim \mathcal{CN}(\mathbf{0},\sigma_s^2\mathbf{I}_{MN})$.
As an initial attempt, we mainly consider the line-of-sight (LoS) path in this paper, i.e., $P=1$. Then, one can adopt CRB to evaluate the estimation performance of the Doppler shift $\nu$ \cite{kay1993fundamentals}, which is given by $ {\mathrm{CRB}}=\left(\frac{1}{\sigma_{\mathrm{s}}^2} \operatorname{tr}\left(\dot{\mathbf{H}}_\mathrm{s,dd}\mathbf{WW}^{H}\dot{\mathbf{H}}_\mathrm{s,dd}^{{H}}\right)\right)^{-1},$
where $\dot{\mathbf{H}}_\mathrm{s,dd}=h_s\left(\mathbf{F}_N \otimes \mathbf{I}_M\right) \mathbf{\Pi}^{l_s} \mathbf{D}_{\nu}\boldsymbol{\Delta}^{k_s}\left(\mathbf{F}_N^{{H}} \otimes \mathbf{I}_M\right)$ and 
$\mathbf{D}_{\nu}$ is given by {\small $\mathbf{D}_{\nu}=\operatorname{diag}\left(j \frac{2 \pi T}{M}[0, \ldots, M N-1]^{\mathrm{T}}\right)$}. The definitions of $h_s$, $l_s$, and $k_s$ are similar to their counterparts in (\ref{timechannel}), respectively.

Based on the above discussion, our goal is to improve the communication performance guaranteed by the BER while satisfying the required CRB and maximum transmission power. Thus, the problem is formulated as 
\begin{align}
     (\mathcal{P}_1) \ \min_{\mathbf{W}} \ \ \ P_e   \quad
\mathrm{s.t.} \quad \mathrm{CRB} \le \gamma_{\nu},
\mathrm{tr}(\mathbf{WW}^H)\le P_0,
\end{align}
where $\gamma_{\nu}$ is the required CRB threshold and $P_0$ is total power.
\vspace{-0.5cm}
\section{ OPTIMAL PRECODER DESIGN}
\label{sec:typestyle}
Problem  ($\mathcal{P}_1$) is difficult to solve directly. As a compromise, in this section, we first establish a lower bound for (\ref{ber}) and adopt it as the new objective function such that the corresponding optimal minimum BER precoder can be designed. According to \cite{9775701}, (\ref{ber}) is convex at moderate-to-high SNRs, i.e., when {\small  $\sigma_c^2\le \eta \left[\left(\kappa\sigma_c^2 \mathbf{I}_{MN}+\mathbf{W}^{{H}} \mathbf{H}_{\mathrm{c,dd}}^{{H}} \mathbf{H}_{\mathrm{c,dd}} \mathbf{W}\right)^{-1}\right]_{m,m}$} with $\eta=4 \beta /\left(\sqrt{(2 \beta \kappa-9)(2 \beta \kappa-1)}+3+2 \beta \kappa\right)$. With this mild condition, a lower bound for (\ref{ber}) can be derived by employing Jensen inequality, which is written as
{\small \begin{align}
 P_e \geq \alpha \operatorname{erfc}\left(\sqrt{\frac{\beta MN}{\sigma_c^2 \phi }-\beta \kappa}\right) \triangleq P_e^\mathrm{lb},
\end{align}}
\hspace{-0.13cm}where {\small $\phi=\operatorname{tr}\left[\left(\kappa\sigma_c^2 \mathbf{I}_{MN}+\mathbf{W}^{{H}} \mathbf{H}_{\mathrm{c,dd}}^{{H}} \mathbf{H}_{\mathrm{c,dd}} \mathbf{W}\right)^{-1}\right]$}. It should be noted that the inequality holds if and only if all entries {\small$\left[\left(\kappa\sigma_c^2 \mathbf{I}_{MN}+\mathbf{W}^{{H}} \mathbf{H}_{\mathrm{c,dd}}^{{H}} \mathbf{H}_{\mathrm{c,dd}} \mathbf{W}\right)^{-1}\right]_{m,m}$ } are identical. For ease of study, we now resort to minimizing the lower bound of $P_e$. Due to the fact that erfc($\cdot$) is monotonically decreasing w.r.t the impact argument, minimizing $P_e^\mathrm{lb}$ is equivalent to the minimization of $\phi$. Again, assuming the signal propagation between the transmitter and the communication user is dominated by a LoS path, such that $\left(\mathbf H_{\mathrm {c,dd}}^{ H}\mathbf H_{\mathrm {c,dd}}\right)=|h_c|^2\mathbf{I}_{MN} $, where $h_c$ is the LoS path gain. Then, we can decompose $\mathbf{W}$ by applying SVD, which yields $\mathbf{W} =\mathbf{U}\mathbf{\Sigma} \mathbf{V}$, where $\mathbf{U}$and $\mathbf{V}$ are both unitary matrices size of ${MN\times MN}$ and $\mathbf{\Sigma} \in \mathbb{C}^{MN\times MN}$is diagonal. Therefore, a suboptimal solution to $ (\mathcal{P}_1) $ can be obtained by solving
{\small \begin{subequations}
  \begin{align}
 (\mathcal{P}2)  \ &\min _{\mathbf{U}, \mathbf{\Sigma}, \mathbf{V}} \operatorname{tr}\left(\left(\kappa \sigma_c^2 \mathbf{I}_{MN}+|h_c|^2\mathbf{\Sigma}^2\right)^{-1}\right) \\
\text { s.t. } & \operatorname{tr}\left(\boldsymbol{\Sigma}^2\right) \le P_0, \\
    &\left(\frac{1}{\sigma_{\mathrm{s}}^2} \operatorname{tr}\left( \mathbf{\Sigma^2}\mathbf{U}^H\dot{\mathbf{H}}_\mathrm{s,dd}^{{H}}\dot{\mathbf{H}}_\mathrm{s,dd} \mathbf{U}\right)\right)^{-1}\le \gamma _\nu, \label{crb} \\
 & {\left[\mathbf{V}^H\left({\kappa \sigma_c^2 \mathbf{I}_{MN}+|h_c|^2\mathbf{\mathbf{\Sigma}^2}}\right)^{-1}\mathbf{V}\right]_{mm} \leq \eta\sigma_c^2}.  \label{convexity}
\end{align}
\end{subequations}}
 $\mathbf{V}$, which only appears in (\ref{convexity}), has no impact on both the BER and CRB, whose function is to ensure the validity of $P_e^\mathrm{lb}$. Thus, without loss of generality, ($\mathcal{P}2$) can be optimized via first solving (\ref{convexity}) for arbitrary $\mathbf{U}$ and $\mathbf{\Sigma}$, and then dealing with the remaining part. As observed, (\ref{convexity}) is equivalent to minimizing the largest diagonal element on the left-hand side of (\ref{convexity}). According to \cite{1223551},  an optimal $\mathbf{V}$, satisfying (\ref{convexity}), exists if and only if $\mathrm{tr}\left( \left(\kappa \sigma_c^2\mathbf{I}_{MN}+|h_c|^2\mathbf{\Sigma}^2 \right)^{-1}\right) \le MN\eta \sigma_c^2$. Then, an available candidate for $\mathbf{V}$ is $\mathbf{\Psi F_{MN}}$, where $\mathbf{\Psi}$ is the eigenvector of $\left(\kappa \sigma_c^2\mathbf{I}_{MN}+|h_c|^2\mathbf{\Sigma}^2 \right)^{-1}$. Furthermore, such a $\mathbf{V}$ actually can guarantee all the diagonal entries are identical, and thus achieve the lower bound $P_e^\mathrm{lb}$. As a result, the remaining problem is as follows: 
{\small \begin{subequations}
    \begin{align}
 (\mathcal{P}3) \ \ \min _{\mathbf{U}, \mathbf{\Gamma}} & \operatorname{tr}\left(\left(\kappa \sigma_c^2 \mathbf{I}_{MN}+|h_c|^2\mathbf{\Gamma}\right)^{-1}\right) \label{trber}\\
\text { s.t. } & \operatorname{tr}\left(\boldsymbol{\Gamma}\right) \le P_0, \ \text{and} \ \operatorname{tr}\left( \mathbf{\Gamma}\mathbf{Z_s\left(U\right)}\right)\ge \tilde{\gamma _\nu}, 
  \end{align}
\end{subequations}}
where $\mathbf{\Gamma}=\mathbf{\Sigma}^2$, $\mathbf{Z_s(U)}=\mathbf{U}^H\mathbf {E_s\Lambda_s E_s}^H \mathbf{U}$ with the term $\mathbf {E_s\Lambda_s\mathbf E_s}^{H}$ being the eigenvalue decomposition of $\mathbf {\dot{H}}_{\mathrm {s,dd}}^H\mathbf {\dot{H}}_{\mathrm {s,dd}}$ and $\tilde{\gamma_{\nu}}=\sigma_s^2/\gamma_{\nu}$. We solve the optimal $\mathbf{\Gamma}$ given $\mathbf{U}$ first and then find the optimal $\mathbf{U}$. For a given $\mathbf{U}$, Problem $(\mathcal{P}3)$ is convex and can be solved by CVX via the interior-point method \cite{grant2014cvx}. However, the interior-point method is with high computational complexity, i.e., $\mathcal{O}(MN)^{3.5}$. To obtain a low-cost solution, we first derive the Lagrangian function given by
{\small \begin{align}
 \nonumber   \mathcal{L}(\mathbf{\Gamma},\lambda,\mu)=&\text{tr}\left(\left(\kappa \sigma_c^2 \mathbf{I}_{MN}+|h_c|^2\mathbf{\Gamma}\right)^{-1}\right)+\lambda(\text{tr}(\mathbf{\Gamma})-P_0)\\ &-\mu \left(\text{tr}(\mathbf{\Gamma}\mathbf{Z_s(U)})-\tilde{\gamma_{\nu}}\right), \label{lag}
\end{align}}
where $\lambda \ge 0$ and $\mu \ge 0$ are the dual variables.
By setting the gradient w.r.t $\mathbf{\Gamma}$ to zero, we obtain the optimal $\mathbf{\Gamma}$ written by
{\small \begin{align}
    \mathbf{\Gamma}=\frac{\left(\lambda\mathbf{I}_{MN}-\mu\mathbf{Z_s(U) }\right)^{-\frac{1}{2}}}{|h_c|}-\frac{\kappa \sigma_c^2\mathbf{I}_{MN}}{|h_c|^2}, \label{gamma}
\end{align}}
in which the optimal $ \mathbf{Z_s(U)}$ must be diagonal since $\mathbf{\Gamma}$ is diagonal. In other words, an optimal choice of $\mathbf{U}$ is one that forces $\mathbf{U}^H\mathbf{E}_s$ to be a permutation matrix, and thus $\mathbf{U}$ can be chosen as $\mathbf{E}_s$\footnote{ Based on (\ref{gamma}), the optimal $\mathbf{U}$ may not be unique in our considered case. For example, $\mathbf{U}$ can be also obtained by switching the order of columns in $\mathbf{E_s}$. Since $\mathbf{U}^H\mathbf{E_s}$ only decides the order of the elements in $\mathbf{\Lambda_s}$, the optimal $\mathbf{\Gamma}$ will be designed following the same order as $\mathbf{\Lambda_s}$ without changing the values of $\mathbf{\Gamma}$, which has no impact on the objective value of (\ref{trber}). It should be noted that, however, the optimal $\mathbf{\Gamma}$ is unique for a given $\mathbf{U}$.}.Then, all that remains is to solve the value of  $\lambda$ and $\mu$. 
Based on the KKT conditions, 
we can immediately observe that $\lambda>0$ since the available power budget should always be exhausted. Furthermore, if the CRB constraint is inactive, i.e., $\mu=0$, we can obtain the optimal $\mathbf{\Gamma}=(P_0/MN) \mathbf{I}_{MN}$.  However, suppose that $\mu >0$, which indicates the CRB constraint is active. In this case, it is intractable to obtain $\lambda$ and $\mu$ directly, we then provide an efficient iterative method to obtain these parameters at low cost. According to (\ref{lag}), we have the following dual problem 
\begin{align}
   (\mathcal{P}4) \ \max _{\lambda \geq 0, \mu \geq 0} \inf _{\mathbf{\Gamma} \succeq \mathbf{0}} \mathcal{L}\left(\mathbf{\Gamma}, \lambda, \mu\right) \triangleq \max _{\lambda \geq 0, \mu \geq 0} \mathcal{D}\left(\lambda, \mu\right), 
\end{align}
where $\mathcal{D}\left(\lambda, \mu\right)$ represents the dual function, which can be obtained by substituting (\ref{gamma}) into (\ref{lag}). Since $ (\mathcal{P}3)$ is convex given $\mathbf{U}$ and satisfies the Slater's conditions, the dual gap between $ (\mathcal{P}3)$ and $ (\mathcal{P}4)$ is zero. To find the optimal $\lambda$ and $\mu$, one can apply the gradient descent method \cite{5406097}, i.e.,
\begin{align}
    \lambda^r=\left[\lambda^{r-1}+\epsilon_1\nabla \mathcal{D}\left(\lambda^{r-1}\right)\right]^{+},\\
     \mu^r=\left[\mu^{r-1}+\epsilon_2\nabla \mathcal{D}\left(\mu^{r-1}\right)\right]^{+},
\end{align}
where $r$ denotes the iteration index, $\epsilon_1$ and $\epsilon_2$ are the corresponding stepsize \cite{boyd2004convex}. Note that the  gradients $\nabla \mathcal{D}\left(\lambda^{r-1}\right)$  and $\nabla \mathcal{D}\left(\mu^{r-1}\right)$ can be calculated straightforwardly. Due to strict page limitations, we omit the detailed derivations here. 
\vspace{-0.5cm}
\section{SIMULATION RESULTS}
In this section, we present the simulation results to verify the effectiveness of our proposed precoder. The system setup is as follows:  we set $M=8$, $N=8$ with $16$-QAM signaling and the carrier frequency is $4$ GHz with sub-carrier spacing being $2 $ kHz. The maximum speed of the mobile user is $v_{\max }=120 \mathrm{~km} / \mathrm{h}$, leading to a maximum Doppler-shift tap $k_{\max }=2$. The maximum delay tap is assumed to be $l_{\max }=4$, and the delay taps belong to $\left[0, l_{\max }\right]$.  The required CRB threshold is set to $3\times 10^{-7}$. For ease of exposition, We employ the following benchmarks for comparisons, i.e., ZF without CRB (ZF-WC) precoding and MMSE without CRB (MMSE-WC) precoding, in which we minimize the BER  without considering CRB constraint. We adopt $||\nabla \mathcal{D}(\lambda)|| \le 10^{-2}$ and $||\nabla \mathcal{D}(\mu)|| \le 10^{-2}$ as the stopping criterion.
\begin{figure}[t]
    \centering
    \includegraphics[width=0.37\textwidth]{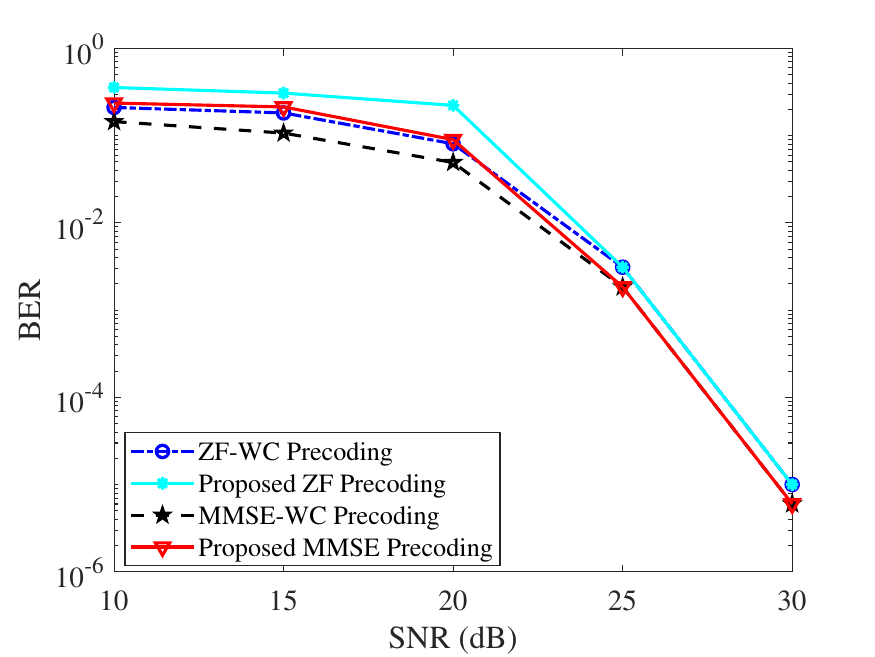}
    \caption{The BER performance versus SNRs under $16$-QAM with CRB threshold $\gamma_{\nu}=3\times 10^{-7}$.  }
    \label{fig1}
\end{figure}
\begin{figure}[t]
    \centering
    \includegraphics[width=0.4\textwidth]{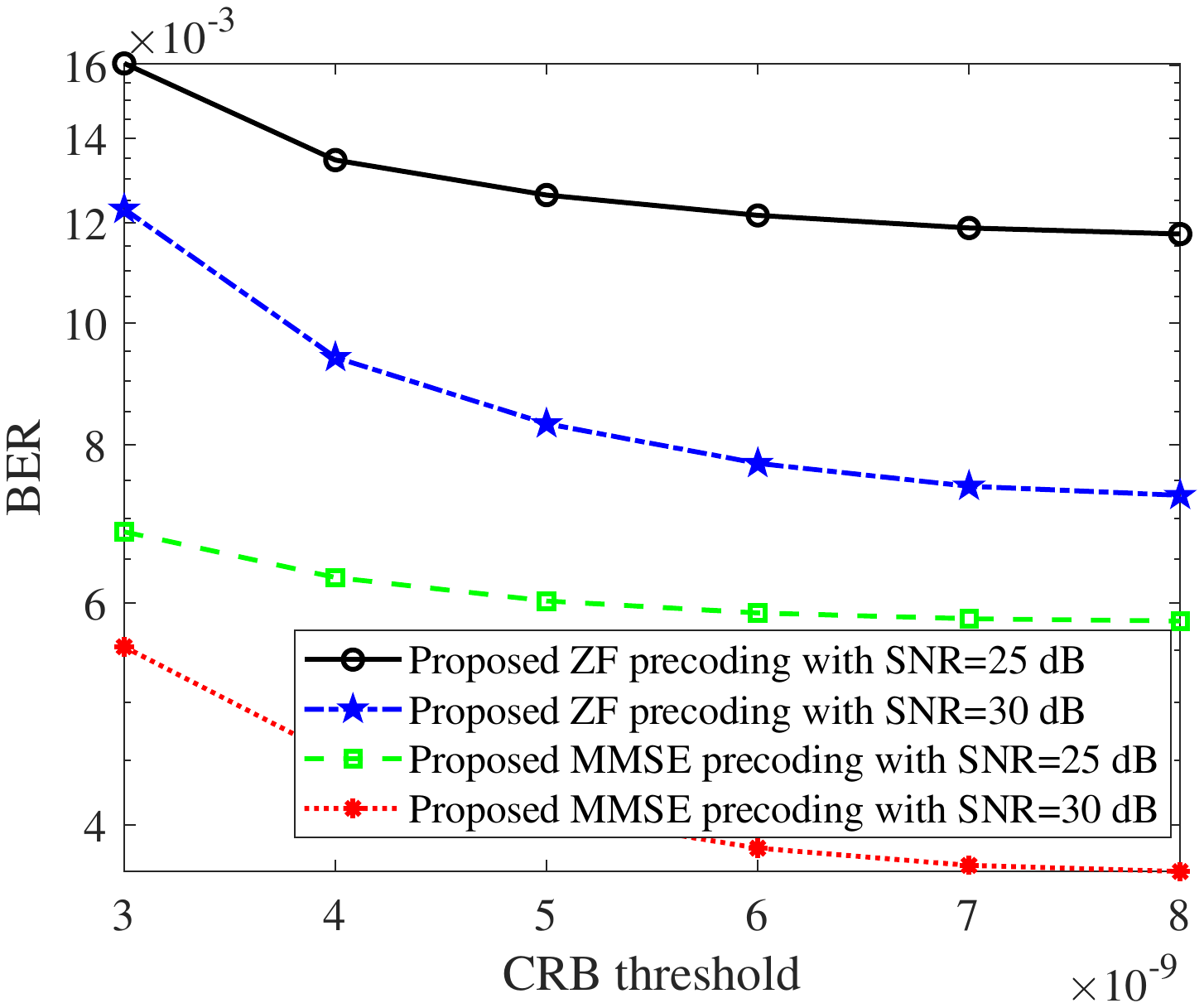}
    \caption{The BER performance with different CRB thresholds. }
    \label{fig2}
    \vspace{-0.3cm}
\end{figure}
Fig. \ref{fig1} illustrates the BER performance with different SNRs. As observed, in the context of SNR ranging from $10$ dB to $20 $ dB, it becomes evident that the proposed ZF and MMSE precoding schemes fail to attain the BER performance achieved by their counterparts, ZF-WC and MMSE-WC schemes.  This disparity can be attributed to that the design of the optimal precoder $\mathbf{\Gamma}$ necessitates a delicate equilibrium between reliable data transmission and satisfying the sensing constraint at lower SNR levels.  Consequently, BER performance exhibits a sacrifice under these conditions. 
As the required SNR increases, the BERs of our proposed ZF/MMSE precoding methods gradually approach that of the ZF-WC/MMSE-WC schemes and eventually coincide with them.  This phenomenon is born of the fact that higher SNRs provide a more generous power budget, allowing for the fulfillment of the pre-set sensing constraints with primary (even only) consideration on the BER. On the other hand,  in our proposed precoding schemes, the optimal $\mathbf{\Gamma}$ has to allocate part of the power to ensure the CRB is satisfied, which indicates that the Doppler shift estimation accuracy will always be better than that of the ZF-WC/MMSE-WC schemes.
To explore the interplay between communication and sensing, we also present the BER performance across varying CRB thresholds, at SNR levels of $25$ dB and $30$ dB, as depicted in Fig. \ref{fig2}.  Notably, as the CRB threshold increases, there is a further improvement in the achievable BER performance as the system has a higher degree of freedom in allocating more power to BER.  Conversely, when the CRB threshold is reduced, the optimal $\mathbf{\Gamma}$ sacrifices a portion of BER performance to ensure that the CRB constraint is satisfied. However, if the CRB thresholds were to be further increased, the CRB constraint would become inactive most of time,  and the BER of our proposed ZF/MMSE precoders become similar to that of ZF-WC/MMSE-WC schemes.
\vspace{-0.4cm}
 \section{CONCLUSION}

 In this paper, we proposed a DD domain dual-functional precoder for OTFS-enabled ISAC systems, where the goal is to minimize the BER while guaranteeing the Doppler estimation performance.  We first formulated the desired problem which minimizes the lower bound of the average BER. To solve this problem at low cost, we proposed an efficient iterative method by focusing on the dual problem. Numerical results demonstrated that our proposed precoder can provide a considerable BER performance and unveil the sensing and communication interplay in the dual-functional precoder design.

\vfill\pagebreak

\bibliographystyle{IEEEbib}
\bibliography{strings.bib}

\end{document}